\documentclass[twocolumn,showpacs,prb,superscriptaddress,preprintnumbers,amsmath,amssymb]{revtex4}
\usepackage[dvips]{graphicx}
\usepackage{dcolumn}
\usepackage{bm}

\newcommand{\cu}{$\rm Cu^{2+}$}

\newcommand{\eu}{$\rm Eu^{3+}$}

\newcommand{\R}{$\rm R^{3+}$}
\newcommand{\la}{$\rm La^{3+}$}
\newcommand{\sr}{$\rm Sr^{2+}$}

\newcommand{\cuNMR}{$^{63}\rm Cu$}

\newcommand{\lsco}{$\rm La_{2-x}Sr_xCuO_4$}
\newcommand{\lfco}{$\rm La_{1.85}Sr_{0.15}CuO_4$}

\newcommand{\lfsco}{$\rm La_{1.65}Eu_{0.2}Sr_{0.15}CuO_4$}
\newcommand{\leseco}{$\rm La_{1.75}Eu_{0.17}Sr_{0.08}CuO_4$}

\newcommand{\lrsco}{$\rm La_{2-x-y}R_{y}Sr_xCuO_4$}
\newcommand{\lessco}{$\rm La_{1.83-x}Eu_{0.17}Sr_xCuO_4$}

\newcommand{\sus}{susceptibility}

\begin{document}

\title{The Local Structure of \mbox{La$_{\rm 1.65}$Eu$_{\rm 0.2}$Sr$_{\rm 0.15}$CuO$_{\rm
4}$} determined  by  $^{63}$Cu NMR Spectroscopy and Van Vleck Paramagnetism of Eu$^{3+}$
Ions}

 \author{B. Simovi\v c}
 \affiliation{Condensed Matter and Thermal Physics, Los Alamos National Laboratory, Los Alamos, NM 87545}
 \author{M. H\"{u}cker}
 \affiliation{Physics Department, Brookhaven National Laboratory, Upton, New York 11973}
 \author{P. C. Hammel}
 \affiliation{Condensed Matter and Thermal Physics, Los Alamos National Laboratory, Los Alamos, NM 87545}
 \author{B. B\"uchner}
 \affiliation{II. Physikalisches Institut, RWTH--Aachen, 52056 Aachen, Germany}
 \author{U. Ammerahl}
 \affiliation{Laboratoire de Physico-Chimie de l'Etat Solide, Universit\'{e} Paris-Sud, 91405, Orsay Cedex, France}
 \author{A. Revcolevschi}
 \affiliation{Laboratoire de Physico-Chimie de l'Etat Solide, Universit\'{e} Paris-Sud, 91405, Orsay Cedex, France}

 \date{\today}

\begin{abstract}
We investigate the local symmetry of the tilting of the CuO$_6$ octahedra  in
La$_{\rm 1.65}$Eu$_{\rm 0.2}$Sr$_{\rm 0.15}$CuO$_{4}$ by means of $^{63}$Cu NMR
spectroscopy and the Van~Vleck susceptibility of the $\rm Eu^{3+}$ ions. The Cu NMR central line lineshape is sensitive to local structure through the coupling of the
$^{63}\rm Cu$ nuclear quadrupole moment to the local electric field gradient.
The $\rm Eu^{3+}$ Van~Vleck susceptibility, as a single ion effect,
locally probes the symmetry of the crystal field at the Eu site. Both
techniques independently provide clear evidence for a change of the local tilt
symmetry at the first order structural transition from the orthorhombic to the
low temperature tetragonal phase, in excellent agreement with the average
structure obtained by diffraction techniques. We conclude that the symmetry of
the average crystal structure accurately represents the symmetry of the
octahedral tilt pattern on a local scale.
\end{abstract}
LA-UR-02-6429

\maketitle

\begin{figure}
 \vspace{0cm}
\begin{center}
\includegraphics[width=1\linewidth]{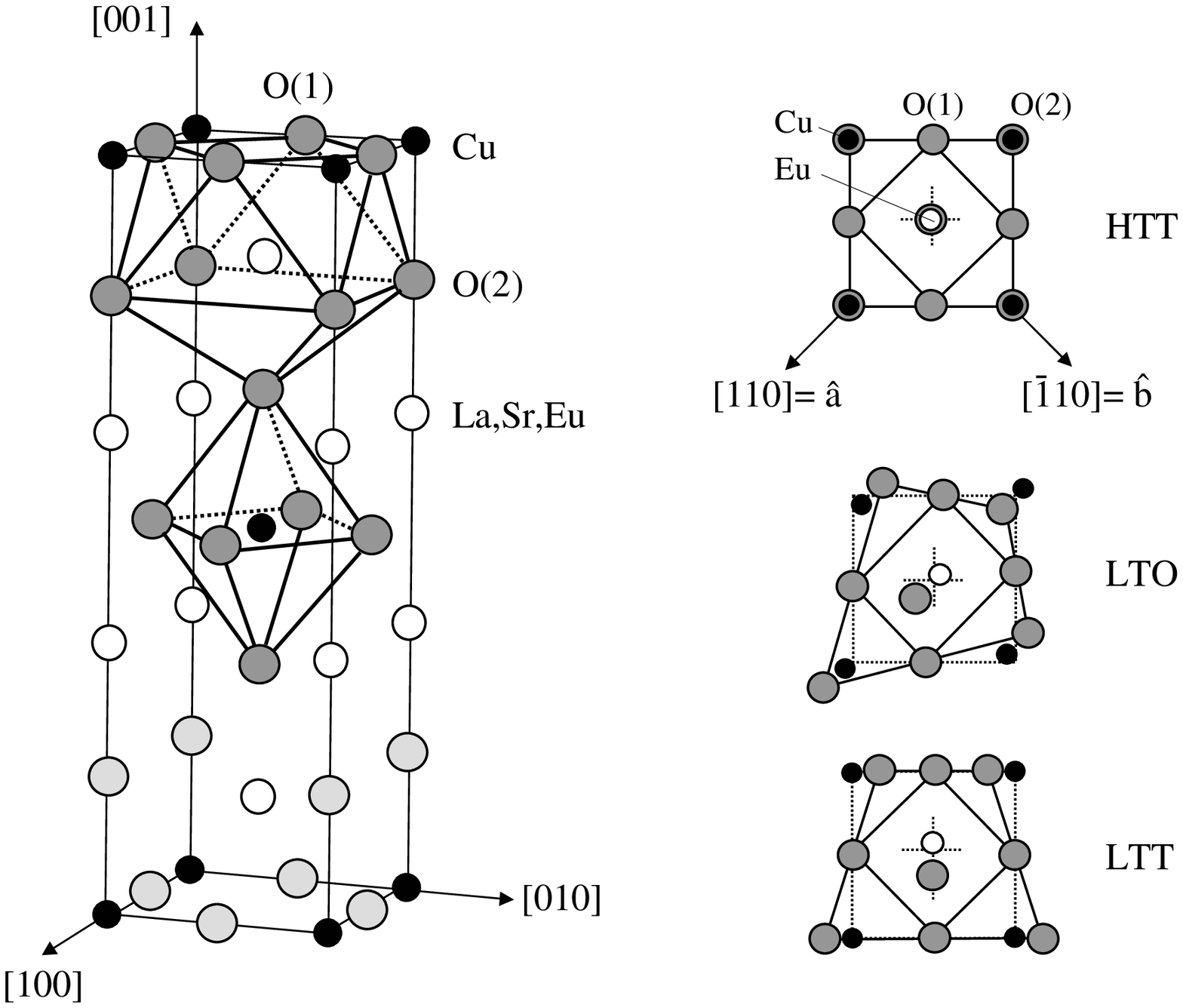}
\caption{ Left: unit cell of \lfsco\ in the HTT phase. A $\rm CuO_6$ octahedron and a
$\rm EuO_9$ polyhedron are accentuated by dark grey oxygens. Right: projections
$\parallel$~$c$ to demonstrate the distortion (exaggerated) of the $\rm EuO_9$ polyhedron
in case of a HTT--, LTO--, and LTT--type $\rm CuO_6$ octahedra tilt pattern. $a$ and $b$
denote the lattice constants of the LTO unit cell.}\label{Fig1}
\end{center}
\end{figure}

Lanthanum cuprate \lsco, the single layer high temperature superconductor, has
been extensively studied over the past few years to understand the mechanism of
superconductivity. Recently, rare earth co-doped lanthanum cuprates \lrsco\ (R
= Eu or Nd) have attracted considerable attention because of the subtle interplay
between charge stripes and superconductivity.~\cite{TranquadaJM:Eviscs} In
this family of compounds it is possible to tune the optimally doped \lsco\
from a superconducting to a magnetic phase where charge is spatially
modulated~\cite{TranquadaJM:Eviscs,WagenerW:MagoL1} by changing the tilt
distortion of the $\rm CuO_2$-layers. These modifications of the structure result
if \la\ is partially substituted by smaller \R\
ions.~\cite{CrawfordMK:Latiec,BuchnerB:Cribds} The suppression of
superconductivity observed in these doping experiments clearly shows an
intimate connection between structure and electronic properties in the high
temperature superconductors.~\cite{CrawfordMK:Latiec,BuchnerB:Cribds} Based on
the structural data obtained from diffraction experiments it has been
argued that both the symmetry and the magnitude of the tilt distortion of the
$\rm CuO_2$ layers are key factors determining the electronic properties in
{\lrsco}\cite{BuchnerB:Cribds, KlaussH-H:antosm}. These neutron and x-ray
diffraction techniques, sensitive to the average structure, show that rare
earth co-doped lanthanum cuprates can exist in three structural phases
dependent upon \sr\ and \R\ doping and upon temperature: the high
temperature tetragonal phase (HTT), the low temperature orthorhombic phase
(LTO) and the low temperature tetragonal phase (LTT). All three phases can be
described by different patterns of rotated $\rm CuO_6$ octahedra. The $\rm
CuO_2$ plane is flat in the HTT phase and the transition to LTO (spacegroup
$Abma$) consists of a tilt of the CuO$_6$ octahedra along the [110] direction
using the notation for the HTT unit cell ($I4/mmm$); see Fig.~\ref{Fig1}. The
tilt angle increases gradually as the temperature is lowered  and the compound
undergoes a first order structural transition to the LTT phase ($P4_2/ncm$) via
a discontinuous change of the tilt direction, which is then oriented alternately along
[100] and [010] in adjacent $\rm CuO_2$
planes.~\cite{AxeJD:Strilc,BuchnerB:Cribds}

Over the past few years local structure models which seem to contradict these
findings have been published. Based on pair distribution
function (PDF) analysis of neutron diffraction and x-ray absorption fine-structure (XAFS)
it was concluded that, locally, the octahedra always tilt with the LTT
symmetry.~\cite{BillingeSJL:LocotL, HaskelD:XAFstl} The long range LTO tilt pattern was
interpreted as the superposition of two out of four equivalent short range LTT patterns
with their tilt axes rotated by $90^{\circ}$, and the flat planes in the HTT phase as the
result of a superposition of all four LTT variants. Furthermore, the amplitude of the
octahedral tilt is thought not to change in this model. This is in sharp contrast to
$^{151}$Eu M\"{o}ssbauer spectroscopy which also probes  the local structure and shows a
temperature dependence of the local tilt angle in \lessco\ which corresponds well with
average structure results.~\cite{FriedrichC:TiltCo,HaskelD:ComTtC, MicklitzH:RepCTt}

It is crucial to clarify this issue in order to gain better insight into the relation
between charge stripes and structural ordering in lanthanum cuprates. In particular,
evidence from  XAFS for locally different Cu-O(1) and Cu-O(2) bond lengths in the LTO
phase of $\rm La_{2-x}Sr_xCuO_4$ have been interpreted as indicating the coexistence of LTO and LTT like
domains.~\cite{BianconiA:Dettll,BianconiA:StrstC, LanzaraA:Tem-de, HaskelD:Doptis} Note
however that similar XAFS experiments performed on  \lrsco\ do not yield evidence for
spatially varying bond lengths.~\cite{NiemoellerT:X-raya} Also, no anomalies related to
the LTO$\rightarrow$LTT transition or the presence of static stripe order could be detected with
this technique.~\cite{NiemoellerT:X-raya} Recently,  PDF results on $\rm
La_{2-x}Sr_xCuO_4$ were discussed in terms of spatially modulated combinations of HTT, LTT, and
LTO tilt patterns which are locally induced by dynamic charge
stripes.~\cite{BozinES:Cha-st} But a detailed analysis of Debye-Waller factors and
diffuse scattering in a neutron diffraction experiment performed on a $\rm
La_{1.85}Sr_{0.15}CuO_4$ single crystal agrees with the average structure, and places low
upper limits on the amplitude of any hidden local structural distortion, including those
induced by stripe correlations.~\cite{BradenM:Anatls}

In this manuscript, we report $^{63}\rm Cu$ NMR and DC magnetic susceptibility
measurements performed on a \lfsco\ single crystal. According to diffraction data \lfsco\
undergoes a second order transition from HTT to LTO at $\sim 350$ K and a first order
transition from LTO to LTT at $T_{\rm LTT}=135$ K. It is known from $\mu\rm SR$
studies that its ground state is magnetic\cite{KlaussH-H:antosm} and a detailed
investigation  of the low frequency spin dynamics by NMR supports the existence of a
glass forming stripe liquid below 30 K.\cite{Simovic}

We will show that the combination of $^{63}\rm Cu$ NMR and DC magnetic susceptibility enables
us to probe the local structure around Cu$^{2+}$ and Eu$^{3+}$ ions as a function of
temperature. The $^{63}$Cu
central line NMR spectrum reflects interaction of the $^{63}\rm Cu$ nuclear
quadrupole moment with the surrounding EFG, and is sensitive to the angle
between the dominant component of the EFG tensor and the applied magnetic field $\bf
H_0$. The magnetic susceptibility is dominated by the Van-Vleck paramagnetism $\chi$(Eu) of the Eu$^{3+}$ ions. Because $\chi$(Eu) is a single
ion effect, its anisotropy reflects the local symmetry of the crystal field at the Eu
site. These two techniques, $^{63}\rm Cu$ NMR and DC magnetic susceptibility, can
discriminate between LTT and LTO and provide clear evidence that the tilt pattern
displays, on a local scale, the symmetry of the macroscopic structure as determined by
diffraction techniques\cite{AxeJD:Strilc}. This leads us to the conclusion that there is
no need to distinguish between the symmetry of the local and average structure in this
compound.

The \lfsco\ single crystal was grown by the traveling solvent floating zone (TSFZ)
method.~\cite{Ammerahl:TZF} Two pieces of $2 \times 2 \times 1\,{\rm mm}^3$ and $5 \times
5 \times 4\,{\rm mm}^3$ were cut from a centimeter size single crystal and oriented by
Laue x-ray back reflection. The $^{63}$Cu ($I = 3/2$) NMR measurements were made on the
central ($m_I = +\frac{1}{2} \leftrightarrow -\frac{1}{2})$ transition of the smaller
crystal which is the same used in the ref\cite{Simovic}. The spectra were obtained by sweeping the static magnetic field $\bf
H_{0} \parallel$[100] at a fixed resonance frequency of 80 MHz. The DC magnetic
susceptibility $\chi=M/H$ with $H=1$~T was measured for all three crystallographic
directions of the LTO unit cell of the larger crystal using a Faraday balance. The
susceptibility data for a \leseco\ crystal are shown for
comparison.

In Fig.~1 we schematically show the unit cell in the HTT phase. The
local charge surrounding  both \cu\ and \eu\ ions will be affected by the tilt of the
CuO$_6$ octahedra. Cu NMR is an excellent tool for
investigating the local structure because atomic displacements locally
alter the charge distribution and hence the EFG.
The EFG tensor is described by its three components; by convention we take $ V_{\rm zz}\geq
V_{\rm yy} \geq V_{\rm xx}$ where $V_{\alpha \alpha}$ is the second derivative of the local
electrostatic potential. The quadrupole frequency $\omega_{Q}$ is proportional
to $\mid V_{\rm zz}\mid$ and the asymmetry parameter is defined as: $\eta=\mid
V_{\rm xx}-V_{\rm yy}\mid/V_{\rm zz}$ which is zero for a tetragonal symmetry.
In the HTT phase, the average tilt angle is $ 0^o$ and as a result the main
component $\bf V_{\rm zz}$ is parallel to the c-axis. We determine from
nuclear quadrupole resonance (NQR) that $\omega_{Q} = 36.4 \rm MHz$ for
\cuNMR\ and point charge calculations show that in the LTO phase $\eta$
remains close to zero.

From a general point of view, the quadrupole coupling is a function of
$\cos^2\xi$ where $\xi$ defines the angle between $\bf V_{\rm zz}$ and ${\bf
H_{0}}$. In a 2$^{nd}$ order perturbation expansion the quadrupole shift of the
central line of a \cuNMR\ (spin 3/2) nucleus is given by\cite{Slichter, Abragam}
\begin{equation}\label{Q}
 \delta \omega = \frac{3 \omega_{Q}^{2}}{16 \omega_{0}}(1-\cos^{2}
\xi)(1-9\cos^{2} \xi)
\end{equation}

\begin{figure}
 \vspace{0cm}
\begin{center}
\includegraphics[width=1\linewidth]{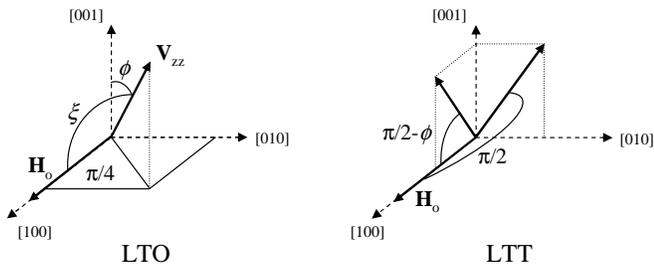}
\caption{Schematic representation of the symmetry of the tilt of
$\bf V_{\rm zz}$  at Cu site in  LTO and LTT structures. The angle
$\phi$ is the tilt angle of the EFG, $\bf H_o$ the external
magnetic field, and $\xi$ the angle between $\bf V_{\rm zz}$ and
$\bf H_o$.} \label{Fig2}
\end{center}
\end{figure}

\noindent
with $\omega_{0}$ the Larmor frequency of the nuclear spin in a field ${\bf H_{0}}$. As
the CuO$_6$ octahedra start tilting in the LTO phase, the main axis of the EFG rotates through
an angle $\phi$ with respect to the c-axis. The connection between the two angles $\phi$
and $\xi$ is shown in Fig.~2; the value of $\xi$ depends on both $\phi$ and the
relative orientation of the applied field ${\bf H_{0}}$ and the tilt direction of
the octahedra. According to the average structure, the CuO$_6$ octahedra tilt along the
diagonal in the LTO phase and alternately along the [100] and [010] directions in
adjacent $\rm CuO_2$ planes in the LTT phase. For this reason, if we apply ${\bf H_{0}}$
along the [100] or [010] directions, we expect to see a single central line in the Cu-NMR
spectra of the LTO phase, and a symmetric splitting of this line in the LTT
phase~\cite{endnote1}reflecting the two different values of $\xi$ in adjacent $\rm CuO_2$ planes:
$\pi/2$ and $\pi/2 -\phi$ (cf. Fig.~2).

Fig.~3 shows $^{63}$Cu NMR central transition swept field spectra as a function of
temperature for ${\bf H_{0}} \|$ [100]. The sharp line at 70.8 kOe is due to the Cu
metal of the coil containing the sample. We see at glance that the \cuNMR\ central line
spectra split symmetrically at temperatures below the first order transition from LTO to LTT
occurring at 135K.  Comparison of swept field spectra obtained at different Larmor
frequencies (80 and 94MHz) shows that the splitting scales as $\omega_{0}^{-1}$, thereby
confirming its quadrupole origin. From  equation~(1), we derive the magnitude of the
splitting as a function of the tilt angle $\phi$:
\begin{equation}\label{Q2}
\Delta \omega = \frac{3 \omega_{Q}^{2}}{16 \omega_{0}}(10\sin^{2} \phi-9\sin^{4} \phi)
\end{equation}

\begin{figure}
\vspace{0.0cm}
\begin{center}
\includegraphics[width=1\linewidth]{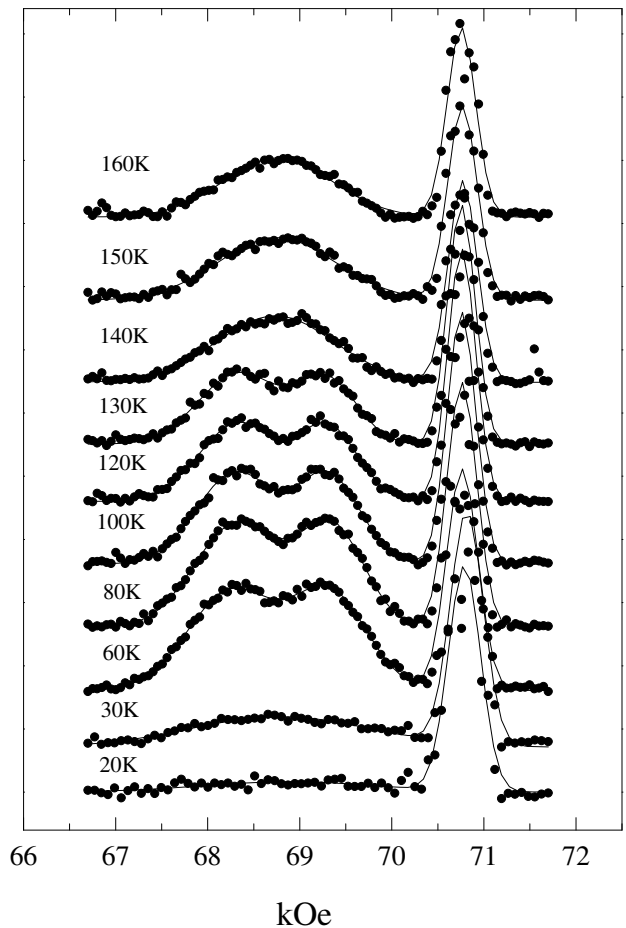}
\caption{$^{63}$Cu NMR spectra at the central transition as a function of temperature:
the field $\bf H_0$ points along [100]. The spectra were obtained by sweeping the field
at a fixed frequency equal to 80 MHz. The narrow line centered at 70.8 kOe comes from the
Cu metal of the NMR coil. } \label{Fig3}
\end{center}
\end{figure}

\noindent
We find $\Delta \omega \approx 1.25 \rm MHz$ from a fit of the
Cu-spectra to two gaussians in the LTT phase (see Fig.~3) and from expression~(2) we obtain $\phi
\approx 12 ^{\circ}$.\cite{endnote2}
In
addition, the average of the split lines corresponds to an angle $\xi \approx
82^{\circ}$. Since $\cos \xi =\cos\pi/4 \sin \phi $ (cf. Fig.~2), we find $\phi \approx
11^{\circ}$ in the LTO phase. Therefore, the amplitude of the tilt angle does not change
significantly at the first order structural transition.

Below 70K, the intensity of the \cuNMR\ NMR signal drops sharply. In Fig.~4 we show the
temperature dependence of the product $I(T)\cdot T$, where $I(T)$ is the integrated
intensity of the central $^{63}$Cu NMR line. The spectra were obtained by
monitoring the intensity of the spin echo versus the static field oriented along the
direction [100] at constant delay $\Delta t \! = \! 10 \mu s$ between the
spin echo excitation pulses.

The Cu signal intensity decreases because, for some Cu nuclei, the transverse magnetization generated by
the radio-frequency excitation decays on a timescale shorter
than the recovery time of the spectrometer, thereby preventing us from observing the full
spin echo signal.~\cite{CurroNJ:Inhlfs} This is a manifestation of a dramatic slowing of the
fluctuations of the Cu 3d magnetic moments below 70K as charge and magnetic order
develop.~\cite{CurroNJ:Inhlfs,Simovic} Since only a fraction of the Cu nuclei
contribute to the full spin echo signal, we cannot make any statement
about the quadrupole splitting at low temperature.
\begin{figure}
 \vspace{0.0cm}
 \begin{center}
 \includegraphics[width=1\linewidth]{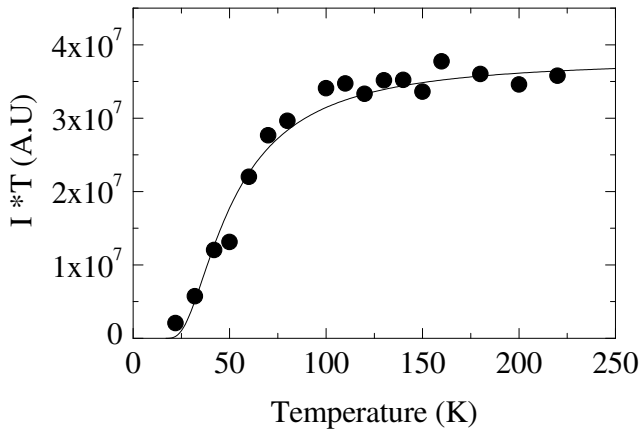}
 \caption{$I \cdot T$ versus $T$ where $I$ is the integrated intensity of
 the central line of the Cu NMR swept field spectra obtained with $\bf H_0$
 along [100] and frequency $\omega_{0}=80$ MHz. The solid line
 is a guide to the eye. } \label{Fig3}
 \end{center}
 \end{figure}

Further insight into the local structure of \lfsco\ is gained from analysis of its DC
magnetic susceptibility $\chi$. The dominant contribution to $\chi$ comes from the
Van~Vleck paramagnetism $\chi(\rm Eu)$ of the \eu\ ions, which is about one order of
magnitude larger than the \sus\ of \lfco\ (cf. Fig.~5). The Van~Vleck paramagnetism is a
single-ion effect and therefore $\chi$ is a suitable tool to investigate the local
symmetry of the $\rm O_9$ oxygen environment surrounding the \eu\ site highlighted in
Fig.~1.

Following Hund's rules, the free \eu\ ion with its $4f^7$ electronic configuration has a
non-magnetic ground state $^7F_0$ ($\rm ^{2S+1}L_{J}$). In an external magnetic field
however the ground state is mixed with the first excited magnetic state, $^7F_{1}$,
thereby leading to a considerable magnetic contribution: the so called zeroth Van
Vleck term. This term is constant at low temperatures~\cite{VanVleck, RettoriC:ESRG3+,
TovarM:Eus24a} but for temperatures T$\gtrsim$ 50 K (cf. Fig.~5) the thermal population of
the excited states $^7F_{J>0}$ leads to further Van~Vleck as well as Curie terms, that
cause the \sus\ to be temperature dependent. In a solid, the $(2J+1)$-fold degeneracy
of the multiplets is lifted according to the symmetry of the crystal field and as a
result, $\chi(\rm Eu)$ becomes anisotropic.~\cite{HundleyMF:Speham,BoothroydAT:Cry-fi,
RettoriC:ESRG3+}

A glance at Fig.~5 shows that this is indeed the case in \lfsco: a large anisotropy is
observed for $H\parallel c$ axis vs. $H \parallel ab$ plane. More important is the
in-plane anisotropy in the LTO phase, which clearly vanishes in the LTT phase. In order
to observe this in-plane anisotropy one must have a partially detwinned \lfsco\ crystal which can be
achieved by application of weak uniaxial pressure. The crystal direction
predominantly containing the shorter $b$-axis has the larger \sus, i.e., $\chi_{b} > \chi_{a}$ in
the LTO phase, using the crystallographic $a$ and $b$ axes of the LTO unit cell with
$a\parallel$~[110] and $b\parallel$~[\={1}10] (cf. Fig.~1).

The increase of the $ab$ anisotropy with decreasing temperature, observed in the LTO
phase in Fig.~5, is a consequence of the continuously increasing tilt angle as was also
observed by $^{151}$Eu M\"{o}ssbauer spectroscopy\cite{FriedrichC:TiltCo}. Close to
$T_{\rm LTT}$ the $ab$ anisotropy reaches a value of $1\times 10^{-4}$~emu/mol. For a sample
with $x=0.08$ (see inset of Fig.~5) the maximum value increases to $3\times
10^{-4}$~emu/mol, which is approximately three times the total $\chi$ of pure \lfco.
The larger $ab$ anisotropy for $x=0.08$ is consistent with the increase of the octahedral
tilt angle as the \sr\ content is reduced. The order of magnitude of the $ab$ anisotropy
in both crystals shows that it originates primarily from the $\rm Eu^{3+}$ Van-Vleck
paramagnetism. The super exchange constant $J$ in the $\rm CuO_2$ planes is very large in
these materials ($\sim 100$ meV) hence the susceptibility of the planes is
small, so unrealistically large distortions would be required to explain
$ab$ anisotropies of the observed magnitude as arising from changes of the Cu spin magnetism.

\begin{figure}[t]
\begin{center}
\includegraphics[width=1\columnwidth, angle=270,clip]{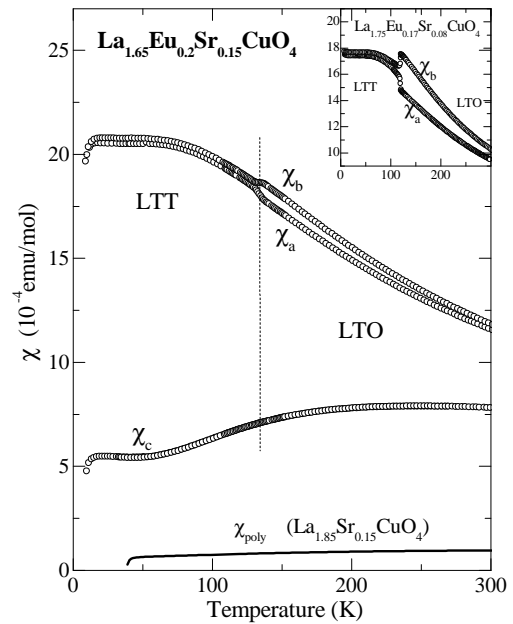}
\caption{Static magnetic susceptibility of $\rm
La_{1.65}Eu_{0.2}Sr_{0.15}CuO_4$ at $H=1$~Tesla for $H\parallel c$ and
$H\parallel ab$ (open circles) and polycrystalline \lfco\ (full line). Inset:
susceptibility of $\rm La_{1.75}Eu_{0.17}Sr_{0.08}CuO_4$ for $H\parallel ab$}
\label{Fig5}
\end{center}
\end{figure}

It is generally assumed that substituting La with smaller rare earth elements such as Nd
or Eu induces strain which may possibly cause a local lattice instability of the LTO
phase towards the LTT phase. From our \sus\ data we can clearly conclude that in the LTO phase
of \lfsco\ the \eu\ site is surrounded by an orthorhombically distorted oxygen
environment. A more quantitative analysis of the $ab$ anisotropy requires a more precise
knowledge of the degree of twinning that is presently lacking. A quantitative analysis of the crystal field
anisotropy between $H\parallel c$ and $H\parallel ab$  will be published elsewhere.

The right side of Fig.~1 shows projections along the unit cell c-axis to
demonstrate the distortion of the $\rm EuO_9$ polyhedron for the HTT, LTO and
LTT local distortions. The measurements of $\chi$ were performed for $H \parallel a$
and $H \parallel b$. It is obvious from these projections that an $ab$ anisotropy can
occur only in the case of LTO-type octahedral tilts and is absent for LTT-type tilts.
Furthermore, the $ab$ anisotropy is zero for all of the other LTT variants
obtained by $90^{\circ}$ rotations about the c-axis the, and for any
combination of the short range LTT domains that are
required to form the LTO and HTT phases in the local structure model first
proposed in Ref~\onlinecite{BillingeSJL:LocotL}. The present results,
therefore, clearly demonstrate that in the LTO phase the $\rm CuO_6$ octahedra tilt
in a LTO-type manner.

To summarize, we have presented investigations of the local structure of a \lfsco\ single
crystal by means of \cuNMR\ NMR and Van-Vleck susceptibility of \eu\ ions as a function
of temperature. Independently, both techniques provide clear evidence that the symmetry
of the local tilt in the LTO and LTT phases is different, in complete agreement with
average structure results\cite{AxeJD:Strilc}. We thus conclude that the symmetry of the
macroscopic structure accurately represents the tilt pattern of the CuO$_6$ octahedra on a
local scale.

The authors thank N.Curro, V. Kataev and H.-H. Klauss  for
valuable discussions. The work at Los Alamos National Laboratory
was performed under the auspices of the US Department of Energy.
The work of M.H. at Brookhaven was supported by the Material
Science Division, US Department of Energy under Contract No.
DE-AC02-98CH10886.

\bibliographystyle{prsty}
\bibliography{bibtex}

\end{document}